\documentclass[letterpaper, 11pt]{article}

\usepackage{authblk}

\usepackage[table]{xcolor}

\usepackage{breakcites}

\usepackage{natbib}
\defcitealias{Mathematica}{Mathematica}

\usepackage{
  amsmath, amsthm, amssymb, mathtools, dsfont, units,   
  graphicx, wrapfig, subfig, float,                     
  listings, color, inconsolata, pythonhighlight,        
  fancyhdr, sectsty, hyperref, enumerate, enumitem, framed }   	

\usepackage[ruled,vlined]{algorithm2e}

\usepackage[left=1.15in, right=1.15in, top=1in, bottom=0.85in, headsep=.2in]{geometry}

\usepackage[bottom]{footmisc}

\allowdisplaybreaks

\sectionfont{\fontsize{14}{15}\selectfont}
\subsectionfont{\normalsize}

\hypersetup{colorlinks=true, linkcolor=blue, urlcolor=blue, citecolor=blue}

\renewenvironment{abstract}
{\small\begin{quote}\noindent \par{\bf \abstractname.}}
{\noindent\end{quote}}

\usepackage{caption}
\newcommand\numberthis{\addtocounter{equation}{1}\tag{\theequation}}

\let\b\mathbf

\let\bg\boldsymbol

\newcommand{\iid}{\stackrel{\smash{\text{iid}}}{\sim}}

\newcommand{\R}{\mathbb{R}}   

\newcommand{\E}{\mathbb{E}}     
\newcommand{\var}{\text{Var}}   

\newcommand{\tth}{\text{th}}	

{\endMakeFramed}

\DeclareMathOperator*{\argmin}{argmin}    
\DeclareMathOperator*{\dv}{d\!}           
\DeclareMathOperator*{\diag}{diag}        

\theoremstyle{definition}

\graphicspath{{Figures/}{../Figures/}}

\begin{document}


\title{\large\bfseries\MakeUppercase{Sparse reconstruction of ordinary}\\ \MakeUppercase{differential equations with inference}}
{\author[1]{\uppercase{Sara Venkatraman, Sumanta Basu, Martin T. Wells}\\ Department of Statistics and Data Science, Cornell University}}
\date{\vspace{-6ex}}
\maketitle


\begin{abstract}
Sparse regression has emerged as a popular technique for learning dynamical systems from temporal data, beginning with the SINDy (Sparse Identification of Nonlinear Dynamics) framework proposed by \citet{brunton_discovering_2016}. Quantifying the uncertainty inherent in differential equations learned from data remains an open problem, thus we propose leveraging recent advances in statistical inference for sparse regression to address this issue. Focusing on systems of ordinary differential equations (ODEs), SINDy assumes that each equation is a parsimonious linear combination of a few candidate functions, such as polynomials, and uses methods such as sequentially-thresholded least squares or the Lasso to identify a small subset of these functions that govern the system's dynamics. We instead employ bias-corrected versions of the Lasso and ridge regression estimators, as well as an empirical Bayes variable selection technique known as SEMMS, to estimate each ODE as a linear combination of terms that are statistically significant. We demonstrate through simulations that this approach allows us to recover the functional terms that correctly describe the dynamics more often than existing methods that do not account for uncertainty.
\end{abstract}

\section{Introduction} \label{sec:intro}

Ordinary differential equations (ODEs) are used in many scientific fields to mathematically describe nonlinearly-evolving temporal phenomena \citep{strogatz2018nonlinear}. As pointed out by \citet{brunton_discovering_2016}, temporal data in such fields are now more abundant than theoretical laws that describe the dynamics of the processes from which the data were sampled. It is therefore of interest to use time series data to statistically learn explicit representations of a system's governing ODEs, which can then be used for forecasting, system control, or identifying system properties such as stability and bifurcations. Estimating ODEs from time series data has become an important task in modern genomics \citep{farina2008embedding, lu_high-dimensional_2011, moris_transition_2016, bucci2016mdsine}, neuroscience \citep{breakspear_dynamic_2017}, epidemiology \citep{hooker_parameterizing_2011}, ecology and population dynamics \citep{ellner_fitting_2002, hogg_exponentially_2020}, and finance \citep{kou_diffusion_2004}, among others.

In this work, we consider a $d$-dimensional dynamical system whose state at time $t$ is $\b x(t) = [x_1(t),...,x_d(t)]^T\in\R^d$, and whose temporal evolution is governed by the differential equation $$\frac{\dv\b x}{\dv t} = \b f(\b x(t))$$ for some unknown function $\b f:\R^d\to\R^d$. Given measurements of the state $\b x$ at $n$ time points, $\{\b x(t_1),...,\b x(t_n)\}$, our objective is to learn $\b f$ in closed form. A well-studied approach to this task is to assume that each component of $\b f$ is a sparse linear combination of several functions. For instance, if the system dimension $d = 2$, we would express $\b f$ as $$\b f(\b x(t)) = [f_1(\b x(t)), f_2(\b x(t))] \text{~~where~~} f_i(\b x(t)) = \sum_{j=1}^m \beta_{ij}f_{ij}\left(x_1(t), x_2(t)\right) \text{~~for~~} i\in\{1,2\},$$ and where $m$ is the total number of functional terms $f_{ij}:\R^d\to \R^d$ that could comprise each component of $\b f$. The non-zero coefficients $\beta_{ij}$ indicate which functions $f_{ij}$ belong in $\b f$. The $f_{ij}$ are often chosen to be polynomials of $\b x(t)$ up to degree $k$. For instance, if $k=2$, we would have: 
\begin{align*}
f_i(\b x(t)) &= \beta_{i0} + \beta_{i1}x_1(t) + \beta_{i2}x_2(t) + \beta_{i3}x_1^2(t)+\beta_{i4}x_2^2(t) + \beta_{i5}x_1(t)x_2(t), ~\text{ for } i\in\{1,2\} \\
&= \begin{bmatrix}1 & x_1(t) & x_2(t) & x_1^2(t) & x_2^2(t) & x_1(t)x_2(t)\end{bmatrix} \begin{bmatrix}\beta_{i0} \\ \beta_{i1} \\ ... \\ \beta_{i5} \end{bmatrix} \numberthis\label{eq:SINDySetup} 
\end{align*}
Sparsity is reflected in the assumption that many of the coefficients $\{\beta_{ij}\}_{j=1}^m$ are zero. One could select any other library of functions $\{f_{ij}\}_{j=1}^m$ according to domain knowledge of the underlying system, such as its known symmetries or periodicity. Noticing that the left-hand side of \eqref{eq:SINDySetup} is a time derivative (which is either observed or, more likely, estimated from the data $\{\b x(t_k)\}_{k=1}^n$) and the right-hand side is linear in $\{\beta_{ij}\}_{j=1}^m$, we can use sparse regression techniques to identify the non-zero coefficients and thus the relevant subset of polynomial terms whose combination describes the dynamics of the underlying system. This procedure, introduced by \citet{brunton_discovering_2016}, is known as the Sparse Identification of Nonlinear Dynamics (SINDy) framework. Extensions of SINDy have been proposed for the recovery and analysis of partial differential equations \citep{schaeffer_learning_2017-1, rudy_data-driven_2017}, chaotic systems \citep{tran_exact_2017}, stochastic differential equations \citep{boninsegna2018sparse}, and multiscale systems \citep{champion_discovery_2019, callaham_learning_2021}, among others. 

However, fewer studies have addressed uncertainty quantification for differential equations learned via SINDy. Substantial advances have been made in developing notions of statistical significance for individual coefficients in a regularized regression model (we refer the reader to \citet{dezeure_high-dimensional_2015} for a comprehensive review), but these developments remain largely unexplored in the context of dynamical system recovery. We propose leveraging these novel statistical inference procedures for learning differential equations, in particular by constructing an estimate $\b{\hat f}$ of $\b f$ comprised of the statistically significant functional terms that describe the dynamics of a given temporal dataset; that is, for each $i\in\{1,...,d\}$, we select the functions $f_{ij}$ whose estimated coefficients $\beta_{ij}$ are significantly different from zero. Our simulations demonstrate that, compared to techniques that simply retain terms with non-zero coefficients, a significance-driven approach yields sparser equations that are closer to those of the true dynamical system that generated the data. That is, it helps prevent the selection of spurious functional terms with small but non-zero coefficients, whose inclusion in the reconstructed function $\b f$ could potentially induce chaotic or otherwise incorrect dynamics. This methodology furthermore allows one to assess the amount of uncertainty and variability in both individual terms of $\b{\hat f}$ as well as $\b{\hat f}$ overall, which is necessary for forecasting the state of the dynamical system at unobserved points in time.  

A few previous studies have developed uncertainty quantification techniques for the recovery of dynamical systems from temporal data. \citet{hirsh_sparsifying_2021} employ Bayesian variable selection using sparsity-promoting priors, thus enabling posterior inference for each functional term that could be included in the reconstructed ODEs. However, this approach relies on a computationally-intensive Markov Chain Monte Carlo (MCMC) sampler. \citet{fasel_ensemble-sindy_2021} and \citet{sashidhar2022bagging} bypass this caveat by using bootstrap aggregation (bagging) to produce empirical inclusion probabilities of each functional term. Lastly, \citet{zhang_robust_2018} use thresholded Bayesian regression to identify relevant terms in $\b f$ rather than the Lasso ($L_1$-regularization; \citet{tibshirani1996regression}) or thresholded least squares, which are the primary regression methods employed by SINDy, and then use the posterior distribution of the included terms to construct error bars for them. 

In this paper, we study the use of three regularized regression techniques for significance-driven recovery of ODE systems: two frequentist approaches, which are bias-corrected versions of the Lasso and ridge regression, and one Bayesian approach, known as Scalable EMpirical Bayes Model Selection (SEMMS). Besides improved accuracy, using these methods for ODE recovery offers two specific advantages. The first is that by virtue of being rooted in rigorous asymptotic theory, this methodology is free of tuning parameters, such as thresholding cutoffs for either coefficients or inclusion probabilities, in contrast to the aforementioned methods. The only parameter the user must ``tune" is the desired significance level (generally denoted $\alpha$, and often set to 0.05), which is a more interpretable parameter than the regularization parameter (generally denoted $\lambda$) that must be selected when using the SINDy framework with its suggested regression techniques, namely the Lasso or sequentially-thresholded least squares. The second advantage of our methodology is computational efficiency, in that the regression techniques we leverage do not require resampling of the given data nor intensive posterior inference. The frequentist bias-correction methods can be run on a single dataset without bootstrapping; the empirical Bayes SEMMS method operates via a generalized alternating maximization (GAM) algorithm that is substantially faster than simulation-based Bayesian methods, such as MCMC, and does not involve estimating the posterior distribution of the regression coefficients. While we focus on the recovery of ODEs in this study, the methodology extends easily to partial differential equations and other types of dynamical systems. 

The remainder of this paper is structured as follows. In Section \ref{sec:sparseReg}, we provide an extended background reviewing the aforementioned assumption that $\b f$ can be written as a linear combination of functional terms and the sparse regression framework that has been developed under this assumption. In Section \ref{sec:biasCorr} we review bias-corrected versions of the Lasso and ridge regression procedures, and in Section \ref{sec:SEMMS} we review SEMMS. Finally, in Section \ref{sec:results} we present the results of applying our significance-based methodology to simulated data generated from well-known ODE systems. We conclude with a discussion summarizing our method and possible extensions, and we include additional simulations in the Appendix.

\section{Background}\label{sec:background}

\subsection{Recovering dynamical systems via sparse regression}\label{sec:sparseReg}

As mentioned in Section \ref{sec:intro}, the assumption underlying the sparse regression framework for learning a differential equation $\dv\b x/\dv t = \b f(\b x(t))$ is that $\b f$ can be written as a linear combination of a small number of possibly nonlinear functions, such as polynomials. This assumption is motivated by the fact that many canonical differential equation models of real-world phenomena can indeed be written this way. In ecology, for example, the {Lotka-Volterra equations} are a model for the population sizes of a prey species $x_1(t)$ and a predator species $x_2(t)$, and are given by: 
$$\frac{\dv x_1}{\dv t}=\alpha x_1-\beta x_1x_2, ~~~~~~~ \frac{\dv x_2}{\dv t} = \delta x_1x_2 -\gamma x_2,$$
where $\alpha,\beta,\gamma,\delta\in\R$ give the rates of increase and decrease for the two species' populations. This system of differential equations can be written as the following matrix equation:
\begin{align*}
{\underbrace{\color{black}\begin{bmatrix}\frac{\dv x_1}{\dv t} & \frac{\dv x_2}{\dv t}\end{bmatrix}}_{\b{\dot x}}} {\color{black}~=~} \underbrace{\color{black}\begin{bmatrix}x_1 & x_2 & x_1x_2\end{bmatrix}}_{\b\Theta(\b x)}\underbrace{\color{black}\begin{bmatrix}\alpha & 0 \\ 0 & -\gamma \\ -\beta & \delta\end{bmatrix}}_{\b B}{\color{black},}
\end{align*}
or more compactly, $\b{\dot x} = \bg\Theta(\b x)\b B$. This matrix representation, in which $\b B$ is sparse (i.e. contains many zeros) suggests that the problem of recovering $\b f$ from time series data $\b x(t_1),...,\b x(t_n)$ is amenable to sparse linear regression. Specifically, we posit that $\b{\dot X} = \bg\Theta(\b X)\b B + \bg\varepsilon$ where $\bg\varepsilon\sim N(\b 0, \b \sigma^2 I_n)$ and, for polynomial degree $k=2$ and state dimension $d=2$ for instance, we would have:
\begin{gather*}
\small\b{\dot X} = \underbrace{\begin{bmatrix}\dot x_1(t_1) & \dot x_2(t_1) \\ ... & ... \\ \dot x_1(t_n) & \dot x_2(t_n)\end{bmatrix}}_{\substack{\text{matrix of time derivatives,}\\ \text{computed numerically}}},  ~~ \b B = \underbrace{\begin{bmatrix}\beta_{0,1} & \beta_{0,2} \\ ... & ... \\ \beta_{5,1} & \beta_{5,2}\end{bmatrix}}_{\substack{\text{unknown sparse}\\ \text{matrix of coeffs.}}},\\
\bg\Theta(\b X) =\underbrace{\begin{bmatrix}1 & x_1(t_1) & x_2(t_1) & x_1(t_1)x_2(t_1) & x_1^2(t_1) & x_2^2(t_1) \\ ... & ... & ... & ... & ... & ... \\ 1 & x_1(t_n) & x_2(t_n) & x_1(t_n)x_2(t_n) & x_1^2(t_n) & x_2^2(t_n) \end{bmatrix}}_{\text{polynomials of observed time series data}}. \numberthis\label{eq:Theta}
\end{gather*}
Each column of the unknown coefficient matrix $\b B$ can be estimated separately via sparse regression methods. The non-zero entries of the estimated matrix $\b{\hat B}$ then indicate which functional terms in the symbolic vector $\bg\Theta(\b x) = [1 ~~ x_1 ~~ x_2 ~~ x_1x_2 ~~ x_1^2 ~~ x_2^2]$ belong in the unknown function $\b f$.

Simulations of this sparse regression procedure for recovering $\dv\b x/\dv t=\b f(\b x(t))$ reveal that the Lasso often selects numerous higher-order polynomial terms that do not actually belong in the true $\b f$. The coefficients of these incorrect terms tend to be small but highly variable across repeated runs, while the coefficients of the correct terms more reliably tend to be larger; this is perhaps unsurprising given the well-known limitations on the the Lasso's predictive performance under highly-correlated design matrices \citep{dalalyan2017prediction}, as is the case in this setting with $\bg\Theta(\b X)$. Overall, such simulations suggest that a notion of statistical significance for these estimated coefficients would provide a more rigorous and accurate assessment of which polynomial terms govern the underlying dynamics of $\b x(t_1),...,\b x(t_n)$, compared to only considering which coefficients are non-zero.

\subsection{Bias-corrected estimates of sparse regression coefficients} \label{sec:biasCorr}

Confidence intervals and $p$-values for common regularized regression procedures, such as the Lasso and ridge regression, have been derived in numerous studies by correcting for bias in the original estimators. We provide a brief description of these methods here and refer to \cite{dezeure_high-dimensional_2015} for additional details and references.

\subsubsection{Bias-corrected Lasso}

The Lasso estimator is defined as the solution to the following optimization problem:
\begin{align*}
\bg{\hat\beta}_\text{Lasso}=\argmin_{\bg\beta}\|\b Y-\b X\bg\beta\|_2^2 + \lambda\|\bg\beta\|_1
\end{align*}
where $\lambda$ is a regularization parameter. Unlike in ordinary least-squares regression, where the least-squares estimator $\bg{\hat\beta}_\text{OLS}=(\b X^T\b X)^{-1}\b X^T\b Y$ is known to have a $N(\bg\beta,\sigma^2(\b X^T\b X)^{-1})$ distribution for $\varepsilon\sim N(0,\sigma^2\b I_n)$, the exact distribution of $\bg{\hat\beta}_\text{Lasso}$ is generally not tractable. Numerous studies in recent years have characterized the  asymptotic behavior of $\bg{\hat\beta}_\text{Lasso}$ with the goal of developing principled statistical inference for its components, such as confidence intervals and $p$-values \citep{zhang2014confidence, van2014asymptotically, javanmard2014confidence}. A particular challenge in doing so is addressing the bias in $\bg{\hat\beta}_\text{Lasso}$ that arises in high dimensions due to the estimation of $\bg\beta\in\R^p$ from data in the lower-dimensional space $\R^n$; $\b X^T\b X$ is rank-deficient when $p>n$. Thus, as one method of recovering the relevant terms in an ODE system, we adopt the approach of calculating a \textit{bias-corrected Lasso estimator} \citep{zhang2014confidence, javanmard2014confidence}. The approximate Gaussianity of this estimator permits the construction of asymptotically valid confidence intervals for its components.

The bias-corrected Lasso estimator we implement is that proposed by \citet{javanmard2014confidence}, for which approximate normality can be derived under any deterministic design matrix $\b X$. To define the corrected estimator, we first define:
$$m_i =\min_{\b m\in\R^p}\b m^T\bg{\hat\Sigma}\b m ~~ \text{subject to} ~ \|\bg{\hat\Sigma}\b m-\b e_i\|_\infty \leq \mu, ~ i = 1,...,p$$
where $\b e_i$ is the $i^\tth$ standard basis vector in $\R^p$, $\bg{\hat\Sigma}=\b X^T\b X/n$ is the empirical covariance matrix of the columns of $\b X$,   and $\mu$, like $\lambda$, is an input to the method. Then, letting $\b M = [m_1, ..., m_p]^T$, the bias-corrected Lasso estimator $\bg{\hat\beta}_\text{BCLasso}$ is defined as: 
\begin{align*}
\bg{\hat\beta}_\text{BCLasso} = \bg{\hat\beta}_\text{Lasso} + \frac{1}{n}\b M\b X^T\left(\b Y - \b X\bg{\hat\beta}_\text{Lasso}\right)
\end{align*}

It is proven in \cite{javanmard2014confidence} that in a high-dimensional regime and under suitable sparsity conditions, $\bg{\hat\beta}_\text{BCLasso}$ is asymptotically normal\footnote{In particular, Theorem 6 states that $\sqrt{n}(\bg{\hat\beta}_\text{BCLasso} - \bg\beta) = \b Z + \bg\Delta$, where $\b Z\sim N(\b 0, \sigma^2 \b M\bg{\hat\Sigma}\b M^T)$ and $\bg\Delta = \sqrt{n}(\b M\bg{\hat\Sigma}-\b I)(\bg\beta - \bg{\hat\beta}_\text{BCLasso})$. That is, the estimation error $\bg{\hat\beta}_\text{BCLasso} - \bg\beta$ can be decomposed into a normally-distributed term with zero mean and a bias term $\bg\Delta/\sqrt{n}$ whose maximum entry is bounded.} with mean $\bg\beta$ and covariance matrix $\sigma^2\b M\bg{\hat\Sigma}\b M^T/n$. This yields the following $100(1-\alpha$)\% confidence interval\footnote{Theorem 15 of \cite{javanmard2014confidence} shows that this interval, denoted $\b J_i(\alpha)$, is asymptotically valid, i.e. that $\lim_{n\to\infty}P\left(\bg\beta_i\in \b J_i(\alpha)\right)=1-\alpha$. } for the $j^\tth$ parameter $\bg\beta_j$:
\begin{align}
\bg{\hat\beta}_{\text{BCLasso},j} \pm \sigma z_{1-\frac{\alpha}{2}}\sqrt{\frac{(\b M\bg{\hat\Sigma}\b M^T)_{jj}}{n}} \label{eq:biasCorrLasso}
\end{align}
where $z_{1-\alpha/2}$ denotes the $(1-\alpha/2)$ quantile of the standard normal distribution, and $\sigma$ is replaced in practice by any consistent estimator $\hat\sigma$ computed from the data.
Correspondingly, we have the following $p$-value for testing the hypothesis $H_{0,j}:\bg\beta_j=0$ against the alternative $H_{A,j}:\bg\beta_j\neq 0$:
\begin{align}
p_j = 2\left[1-\Phi\left(\frac{\sqrt{n}|\bg{\hat\beta}_{\text{BCLasso},j}|}{\hat\sigma\sqrt{(\b M\bg{\hat\Sigma}\b M^T)_{jj}}}\right)\right], \label{eq:pValueLasso}
\end{align}
where $\Phi(x)=\frac{1}{\sqrt{2\pi}}\int_{-\infty}^x e^{-t^2/2}\dv t$ denotes the standard normal CDF. 

Theorem 16 in \citet{javanmard2014confidence} shows that when $p$-values are computed as in \eqref{eq:pValueLasso} and $H_{0,j}$ is rejected for $p_j\leq \alpha$, and we set $\mu=a\sqrt{\log p/n}$ and $\lambda=\sigma\sqrt{c^2\log p/n}$ for large constants $a$, $c$, the type-I error rate is controlled at level $\alpha$ as $n\to\infty$ for any fixed sequence of tests of the hypotheses $\{H_{0,j}\}_{j\in\{1,...,n\}}$. While the original studies proposing this estimator have assumed that the rows of $\b X$ are independently and identically distributed, more recent studies have established asymptotic normality of $\bg{\hat\beta}_\text{BCLasso}$ with this assumption relaxed in the context of autoregressive time series problems \citep{basu2019system}.

For our problem of sparse ODE recovery, we calculate the  bias-corrected Lasso estimator \eqref{eq:biasCorrLasso} and corresponding $p$-values \eqref{eq:pValueLasso} for each component using the R code provided at \url{https://web.stanford.edu/~montanar/sslasso/}. 

\subsubsection{Bias-corrected ridge regression}

An alternative to the bias-corrected Lasso estimator is the bias-corrected ridge estimator \citep{shao_estimation_2012, buhlmann2013statistical, javanmard2014confidence}. Standard ridge regression provides the following estimate of $\bg\beta$: 
\begin{align*}
\bg{\hat\beta}_\text{Ridge} = \argmin_{\bg\beta}\|\b Y - \b X\bg\beta\|_2^2 + \lambda\|\bg\beta\|_2^2 = \left(\b X^T\b X+\lambda\b I_p\right)^{-1}\b X^T\b Y
\end{align*}
where $\lambda$ is a regularization parameter. Since $\bg\beta$ is generally not identifiable in the $p>n$ setting, \citet{shao_estimation_2012} argue that it suffices to instead estimate the projection of $\bg\beta$ onto $\mathcal R(\b X)$, the row space of $\b X$. This projection, denoted $\bg\theta$, is given by $\bg\theta=\b P\bg\beta$, where $\b P = \b X^T(\b X\b X^T)^-\b X$ is the projection operator onto $\mathcal R(\b X)$ and $(\b X\b X^T)^-$ denotes a generalized inverse of $\b X\b X^T$. We can write the $j^\tth$ component of $\bg\theta$ as
\begin{align*}
\bg\theta_j &= \sum_{k=1}^p \b P_{jk}\bg\beta_k = \b P_{jj}\bg\beta_j + \sum_{k\neq j}^p \b P_{jk}\bg\beta_k
\end{align*}
hence:
\begin{align}
\frac{\bg\theta_j}{\b P_{jj}} = \bg\beta_j + \sum_{k\neq j}^p\frac{\b P_{jk}}{\b P_{jj}}\bg\beta_k. \label{eq:projectedRidge}
\end{align}
As observed in \cite{buhlmann2013statistical}, \eqref{eq:projectedRidge} suggests that the discrepancy between the projection $\bg\theta$ and the parameter of interest $\bg\beta$ is $\sum_{k\neq j}^p\b P_{jk}\bg\beta_k/\b P_{jj}$. This quantity can be estimated by $\sum_{k\neq j}^p \b P_{jk}\bg{\hat\beta}_k/\b P_{jj}$, where $\bg{\hat\beta}_k$ is the $k^\tth$ component of an initial estimator, such as the ordinary Lasso. In subtracting this estimated discrepancy from $\bg{\hat\beta}_\text{Ridge}$, we arrive at the \textit{bias-corrected ridge estimator} $\bg{\hat\beta}_\text{BCRidge}$, whose components are given by:
\begin{align*}
\bg{\hat\beta}_{\text{BCRidge},j} = \frac{\bg{\hat\beta}_{\text{Ridge},j}}{\b P_{jj}} - \sum_{k\neq j}^p \frac{\b P_{jk}}{\b P_{jj}}\bg{\hat\beta}_k, ~~ j=1,...,p.
\end{align*}
Assuming normality of $\bg\varepsilon$, with $\sigma_{\bg\varepsilon}=\var(\bg\varepsilon_i)$ for $i=1,...,n$, the quantity $\frac{1}{\sigma_{\bg\varepsilon}}(\bg{\hat\beta}_\text{Ridge}-\bg\theta)$ is approximately normal for $\lambda\searrow 0^+$ \citep{dezeure_high-dimensional_2015}. The same authors use this fact to derive the following the $p$-value corresponding to the test of  $H_{0,j}:\bg\beta_j = 0$ against $H_{A,j}:\bg\beta_j\neq 0$:
\begin{gather*}
p_j = 2\left[1-\Phi\left(\frac{1}{\sigma_{\bg\varepsilon}\sqrt{\bg\Omega_{jj}}}|\b P_{jj}|\left(|\bg{\hat\beta}_{\text{BCRidge},j}|-\Delta_j\right)_+\right)\right] \numberthis\label{eq:ridgePValue}\\
\text{where } \bg\Omega=\left(\bg{\hat\Sigma}+\lambda \b I_p\right)^{-1}\bg{\hat\Sigma}\left(\bg{\hat\Sigma}+\lambda\b I_p\right)^{-1}  \text{ and } \Delta_j = \max_{k\neq j}\left|\frac{\b P_{jk}}{\b P_{jj}}\right|\left(\frac{\log p}{n}\right)^{1/2-\xi}.
\end{gather*}
The above includes the unknown error variance $\sigma_{\bg\varepsilon}$, which in practice can be estimated consistently via the scaled Lasso \citep{sun2012scaled}, as well as $\xi$, which is typically\footnote{The parameter $\xi$ is derived from bounds on the entries of the vector $\bg\Delta$. In particular, Appendix A.1 of \cite{dezeure_high-dimensional_2015} discusses the fact that for normally-distributed errors $\bg\varepsilon$ and with high probability, we have $$|\bg\Delta_j| = O\left(\|\bg\beta\|_0\sqrt{\frac{\log p}{n}}\max_{k\neq j}\left|\frac{\b P_{jk}}{\b P_{jj}}\right|\right).$$ Therefore, if we have $\|\bg\beta\|_0=O((n/\log p)^\xi)$ for some $\xi\in(0,1/2)$, we can bound $|\bg\Delta_j|$ further as: $$|\bg\Delta_j| \leq O\left(\left(\frac{\log p}{n}\right)^{1/2-\xi}\max_{k\neq j}\left|\frac{\b P_{jk}}{\b P_{jj}}\right|\right).$$ That is, for smaller $\xi$, meaning a more sparse $\bg\beta$, we achieve smaller errors in the bias correction.} set to 0.05.

We refer the reader to \citet{dezeure_high-dimensional_2015} for a precise statement of the assumptions needed to derive $p$-values for the bias-corrected ridge and Lasso estimators. However, we note that a sufficient condition for deriving \eqref{eq:ridgePValue} is that $\bg\beta$ is sparse; specifically, it is required that $\|\bg\beta\|_0 = O((n/\log p)^\xi)$.

For our problem of sparse ODE recovery, we calculate the bias-corrected ridge estimator and corresponding $p$-values for each component using the \texttt{hdi} R package \citep{dezeure_high-dimensional_2015}. In particular, we use the Bonferroni-Holm adjusted $p$-values returned by the \texttt{ridge.proj} function to assess the statistical significance of each candidate function in the ODE. We note that a computational advantage of the bias-corrected ridge estimator over its Lasso counterpart is that it can be directly calculated in closed form, i.e. without numerical optimization.

\subsection{Scalable empirical Bayes model selection (SEMMS)} \label{sec:SEMMS}

In addition to the aforementioned inference methods developed for the Lasso and ridge regression, we also employ an empirical Bayes approach to variable selection in generalized linear models developed by \citet{bar_scalable_2020}. This approach, called Scalable EMpirical Bayes Model Selection (SEMMS), is aligned with our use of sparse regression methods in the context of ODE recovery because it assumes that a small but unknown set of candidate predictors (functional terms) have a non-zero effect on the response (the time derivative). SEMMS expresses the variable selection problem as a classification problem in that each candidate term is to be classified as having a positive, negative, or null (zero) effect. In particular, the relationship between the predictors and the response is written as:
\begin{align*}
\E(\b Y) = \b X\bg\Gamma\b u
\end{align*}
where $\b X$ is an $n\times p$ design matrix as before. Here, $\bg\Gamma = \diag\{\gamma_1,...,\gamma_2,...,\gamma_p\}$, with each $\gamma_k$ indicating whether the $k^\tth$ predictor has a negative, null, or positive effect on the response with probabilities $p_L, p_0, p_R$ respectively; that is, $\gamma_k \iid \text{Multinomial}(-1,0,1; p_L, p_0, p_R)$. Finally, $\b u = [u_1,...,u_p]^T$ is a vector of coefficients with $u_k\sim N(\mu, \sigma^2)$ independently of $\gamma_k$. Thus, SEMMS places a three-component normal mixture prior on the coefficients $\bg\Gamma\b u$, with the latent variables $\gamma_1,...,\gamma_p$ indicating which variables have a non-null effect and should be included in the model. SEMMS estimates the parameters using a Generalized Alternating Minimization (GAM) algorithm, a computationally efficient and convergent variant of the EM algorithm. 

For our problem of sparse ODE recovery, we use SEMMS to identify which functional terms in the design matrix $\bg\Theta(\b X)$ have a non-zero relationship to each column of time derivatives in $\b{\dot X}$. We then fit a standard linear regression model of $\dot x_k(t)$ as a function of the selected terms, for each $k\in\{1,...,d\}$, where $d$ is the dimension of the ODE system. Inference is performed via $t$-tests and confidence intervals for each term.

\section{Results} \label{sec:results}

\begin{figure}[H]
\centering\includegraphics[width=\textwidth]{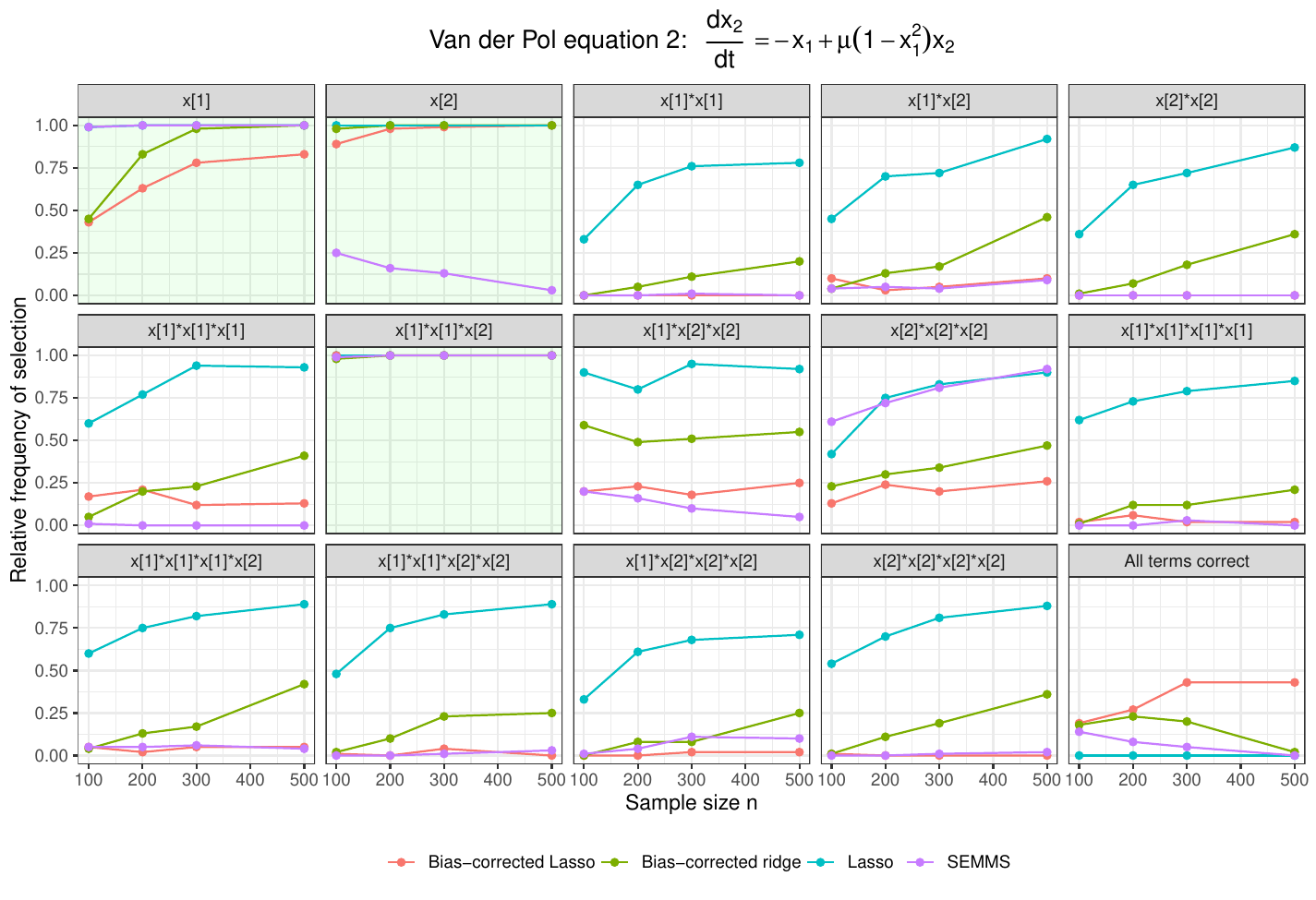}
\caption{Empirical probabilities of selection for each polynomial term up to fourth degree that could potentially be included in the reconstruction of the Van der Pol equation, $\frac{\dv x_2}{\dv t} = -x_1+\mu(1-x_1^2)x_2$ with $\mu=2$, for varying \textit{sample sizes}. The sample size $n$ refers to the number of time steps taken between $t=0$ and $t=15$ when numerically solving the Van der Pol equation with via RK4. Noisy data is obtained by adding i.i.d. $N(0, 0.25)$ random variates to the numerical solution; this was repeated 100 times to obtain the empirical probabilities. Plots corresponding to the correct terms $x_1$, $x_2$, and $x_1^2x_2$, are highlighted in green. The standard Lasso frequently selects incorrect terms even at larger sample sizes, whereas the bias-corrected Lasso and ridge methods as well as SEMMS do not pick up these terms as often. The bottom right plot depicts the empirical probabilities of only these three terms appearing in the recovered equation, i.e. it indicates how often the correct equation was exactly identified.}
\label{fig:VdPselN}
\end{figure}

\begin{figure}[H]
\centering
\includegraphics[width=\textwidth]{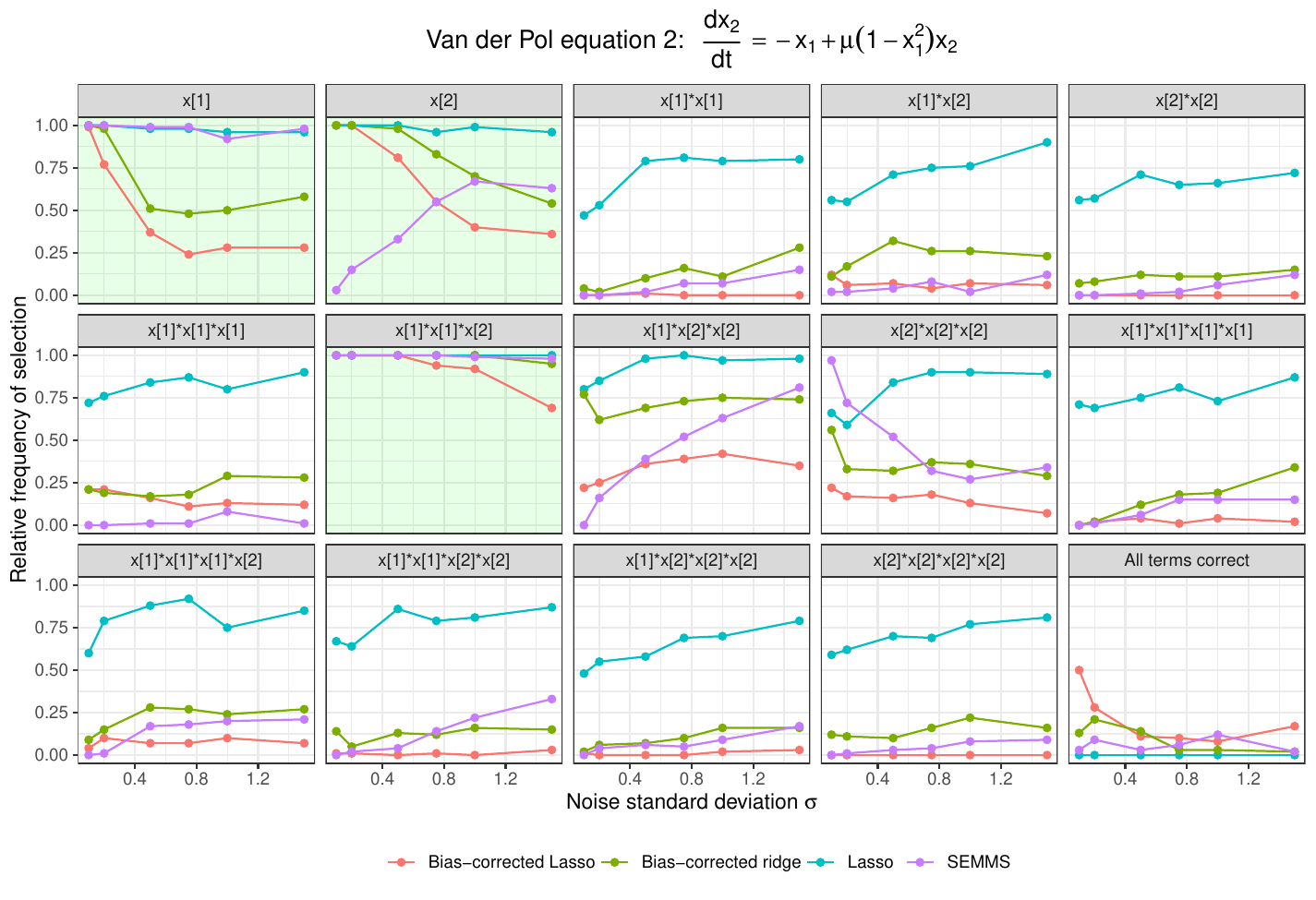}		
\caption{Empirical probabilities of selection for each polynomial term up to fourth degree that could potentially be included in the reconstruction of the Van der Pol equation with $\mu=2$ for varying \textit{noise levels}. The noise level $\sigma$ refers to the standard deviation of the i.i.d. $N(0,\sigma)$ random variates that are added to the numerical solution of the Van der Pol equation obtained via RK4. While incorrect terms are somewhat more likely to be selected by any method under larger amounts of noise, reconstruction via the bias-corrected Lasso and ridge estimators as well as SEMMS appears more robust to noise: incorrect terms are generally selected less than 25\% of the time even at the highest noise levels.}
\label{fig:VdPselS}
\end{figure}

We now apply our significance-driven sparse regression methodology to data simulated from the Van der Pol oscillator. This is a second-order differential equation given by: $$\frac{\text d ^2 x}{\text d t^2}-\mu(1-x^2)\frac{\text d x}{\text d t} + x = 0,$$
where $\mu\in\mathbb R$. We begin by writing this as the following system of two first-order differential equations:
\begin{align}
\frac{\text d x_1}{\text d t} &= x_2, \label{eq:VdP1} \\
\frac{\text d x_2}{\text d t} &= -x_1 + \mu\left(1-x_1^2\right)x_2. \label{eq:VdP2}
\end{align}
To generate data from this system with $\mu=2$, we numerically solve these equations using the Runge-Kutta (RK4) method with an initial condition of $\b x_0 = [x_1(t_0) ~~ x_2(t_0)] = [1,0]$ and step size of $h$ over the time domain $t\in[0,15]$, resulting in $n=15/h$ time steps. This yields the $(n+1)\times 2$ state matrix
$$\b X = \begin{bmatrix} x_1(t_0) & x_2(t_0) \\ x_1(t_1) & x_2(t_1) \\ ... & ... \\ x_1(t_n) & x_2(t_n)\end{bmatrix}.$$
We add i.i.d. Gaussian noise to $\b X$ and compute cubic smoothing splines of each of its columns in order to obtain smoothed approximations to each component of $\b x(t)$. We use this smoothed data to generate the ``feature matrix" $\bg\Theta(\b X)$, consisting of up to fourth-order polynomials and products of the columns of $\b X$, as well as the matrix of time derivatives $\b{\dot X}$, both defined in \eqref{eq:Theta}. We then use these matrices to recover the governing equations \eqref{eq:VdP1} and \eqref{eq:VdP2} using the bias-corrected Lasso and ridge estimators or SEMMS; that is, we regress each column of $\b{\dot X}$ on $\bg\Theta(\b X)$ using these methods, and take the significant coefficients to indicate which polynomial or product terms belong in the equations. Observe that the terms we aim to correctly identify are $x_2$ for \eqref{eq:VdP1} and $x_1$, $x_2$, and $x_1^2x_2$ for \eqref{eq:VdP2}. Using the following metrics and graphics, we compare these methods to the standard Lasso and its set of non-zero coefficient estimates to show that the significance-driven approach reconstructs the equations more accurately:

\begin{enumerate}
\item \textit{Uncertainty in the model coefficients estimated for a single dataset}: Using the bias-corrected Lasso and ridge estimators, we consider the $p$-values associated with the estimated coefficients of each possible polynomial or product term. As explained in Section \ref{sec:biasCorr}, these $p$-values come from tests of the null hypothesis $H_0: \beta_j=0$ against the alternative hypothesis $H_A:\beta_j\neq 0$, where $\beta_j$ is the coefficient of the $j^\tth$ polynomial term; we define a statistically significant result as a $p$-value below 0.05. For SEMMS, we instead consider the posterior probability of $\beta_j$ being 0. We plot the coefficient point estimates and their confidence intervals in Figure \ref{fig:VdPcoeffs}. 

\item \textit{Success rate over multiple datasets}: We next generate many noisy datasets from the ODE system. For each polynomial term that could be included in the recovered equations, we compute the fraction of these noisy datasets in which the term was selected. The overall success rate is defined as the fraction of datasets in which the exact set of zero and non-zero coefficients was correctly identified. We visualize these metrics in Figures \ref{fig:VdPselN} and \ref{fig:VdPselS} by plotting these fractions as they vary with sample size $n$ and the standard deviation $\sigma$ of the Gaussian noise. 
\end{enumerate}

We refer to polynomial terms that are correctly selected by any of the aforementioned sparse regression methods as \textit{true positives}. Selected terms are statistically significant at a 0.05 level under the bias-corrected Lasso or ridge estimators, have a high posterior probability of being non-zero under SEMMS, or have a non-zero coefficient estimate under the Lasso. We refer to incorrectly selected polynomial terms as \textit{false positives}. Figures \ref{fig:VdPcoeffs}--\ref{fig:VdPselS} demonstrate that the bias-corrected Lasso and ridge estimators as well as SEMMS result in many fewer false positives than the Lasso, which often selects high-order polynomial terms with very small but non-zero coefficients. 

\begin{figure}[H]
\centering\includegraphics[width=.85\textwidth]{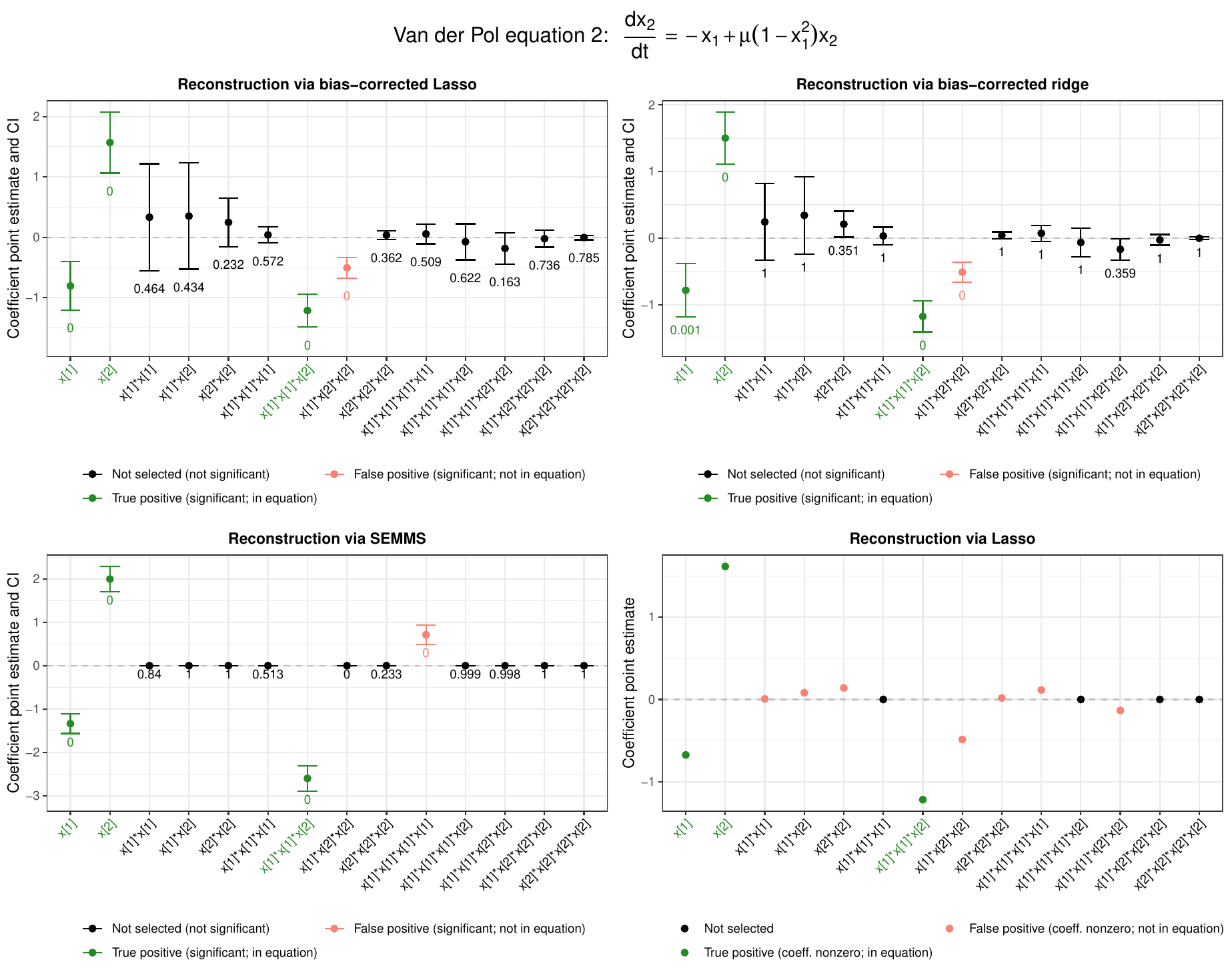}		
\end{figure}
\begin{figure}[H]
\caption{Estimates and error bars for the coefficients of each polynomial term up to fourth degree that could potentially be included in the reconstruction of the Van der Pol equation, $\frac{\dv x_2}{\dv t}=-x_1+\mu(1-x_1^2)x_2$ with $\mu=2$. Coefficient estimates are provided by four sparse regression methods: the bias-corrected Lasso and ridge estimators (upper left and right, respectively), which provide confidence intervals and $p$-values for each term; SEMMS (bottom left); and the Lasso (bottom right). We generate data by adding i.i.d. $N(0, 0.25m)$ noise to the RK4 numerical solution $\b X$ of the Van der Pol system, where $m=\max_{i,j} |\b X_{ij}|$. Beneath the error bars for the bias-corrected Lasso and ridge estimators, we display the corresponding $p$-values. Beneath the bars for SEMMS, we display the posterior probability of the coefficient being zero. }
\label{fig:VdPcoeffs}
\end{figure}

In Figure \ref{fig:VdPselN}, we observe that even at higher temporal resolutions (i.e., higher sample sizes), the standard Lasso frequently selects incorrect polynomial terms, whereas the bias-corrected sparse regression methods select them much less frequently across all temporal resolutions. In Figure \ref{fig:VdPselS}, we observe that the bias-corrected regression estimators are more robust to noise than the standard Lasso in that they continue to exclude incorrect terms even when more noise is added to the data.

Appendix \ref{appsec:VdP} presents additional simulations on the Van der Pol equation, including the empirical distributions of each possible polynomial term's coefficient point estimates over many simulated datasets. In Appendix \ref{appsec:spiral}, we present similar simulations to those shown in this section for a second ODE system.

\subsection{Comparison to Ensemble-SINDy}

We now apply Ensemble-SINDy \citep{fasel_ensemble-sindy_2021} to our simulated data from the Van der Pol oscillator. Ensemble-SINDy (E-SINDy) integrates ensembling techniques into the SINDy framework to improve robustness against noise. Specifically, the method generates $q$ bootstrap samples of the available noisy data in $\bg\Theta(\b X)$ in Equation \eqref{eq:Theta} and applies SINDy to each one, resulting in $q$ learned models of the ODE or PDE system. A level of uncertainty can be associated with each candidate term in $\bg\Theta(\b X)$ via an empirical inclusion probability, i.e. the empirical probability of each candidate term having a non-zero coefficient across the $q$ models. An aggregated model can be constructed by taking the mean or the median of the $q$ coefficients for each possible term, and excluding those with low inclusion probabilities below some threshold.

\begin{figure}[H]
\centering\includegraphics[width=11cm]{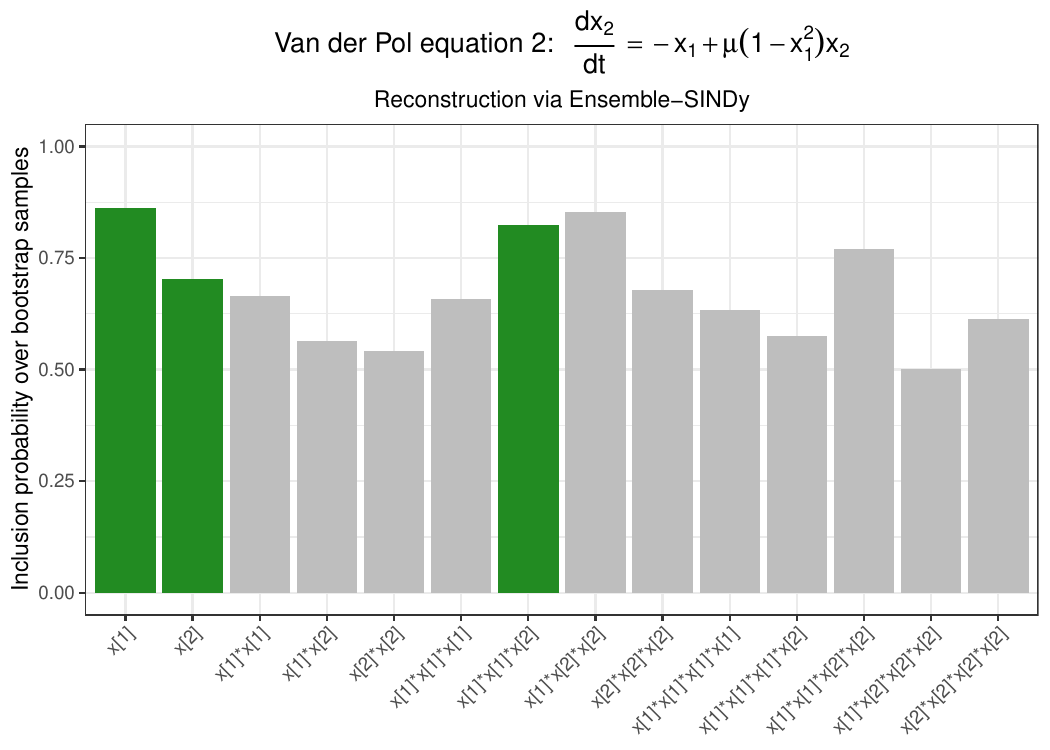}
\label{fig:ESINDy}
\caption{E-SINDy empirical inclusion probabilities of each polynomial term up to fourth degree that could potentially be included in the reconstruction of the Van der Pol equation with $\mu=2$. Data was generated as described in Section \ref{sec:results}. We then generated $q=500$ bootstrap samples from this data, and used SINDy with cross-validated Lasso to estimate coefficients for each term in each bootstrap sample.}
\end{figure}

While the inclusion probabilities for the correct terms in the equation, i.e. $x_1$, $x_2$, and $x_1^2x_2$, are higher than those of the incorrect terms, the distribution of inclusion probabilities does not clearly indicate which terms have the highest likelihood of appearing in the equation. This makes it challenging to select a reasonable threshold for these probabilities that would result in the most correct model. On the other hand, the bias-corrected Lasso and ridge estimators as well as SEMMS automatically select the most appropriate terms without additional thresholding parameters. Prior studies have indicated that the bootstrap is not consistent for the Lasso \citep{chatterjee2011bootstrapping}, but that bootstrapping bias-corrected versions of sparse regression estimators may be more promising \citep{dezeure2017high}. 

\section{Conclusion and future directions}

This study presents a procedure to recover the governing equations of a nonlinear dynamical system from noisy time series data in a way that allows for the uncertainty in the recovered equations to be quantified through $p$-values and confidence intervals; these uncertainty measures are derived from recent advances in asymptotic theory for regularized regression models. The regression techniques considered in this work are bias-corrected versions of the Lasso and ridge estimators, as well as SEMMS, an empirical Bayes variable selection technique for generalized linear models. In particular, our work builds on the well-studied foundation of considering differential equations to lie in a space of basis functions, such as polynomials, and using sparse regression methods to identify a parsimonious subset of these functions whose combination accurately describes the underlying dynamics of the given data. Previous studies have considered uncertainty quantification for this task in the form of sparse Bayesian regression, which allows for the posterior distribution of each possible functional term to be estimated but is computationally intensive, as well as ensemble techniques, which  allow for inclusion probabilities to be estimated over many bootstrap samples but may be ineffective for regularized regression.

Prior work on the use of sparse regression for dynamical system recovery has primarily involved the use of the Lasso or sequentially thresholded least squares regression, though we have found that this tends to result in many spurious terms with small coefficients being included in the recovered equations. By assessing the variance and statistical significance of each candidate term, we can more rigorously justify the inclusion or exclusion of each to produce a more parsimonious and interpretable model. To illustrate our methodology, we conduct simulations in which we generate noisy data from a well-known system of differential equations and attempt to recover these equations using the Lasso, the bias-corrected Lasso and ridge estimators, and SEMMS. Our results show that these last three methods are able to identify the correct functional terms under the criterion of statistical significance (while excluding the others) more often than the Lasso alone, even in the more complex regimes of small sample sizes or increased noise. This demonstrates that statistical inference can substantially improve the robustness of data-driven reconstruction of dynamical systems, and that point estimation alone may not always suffice for this task.

Although we primarily consider systems of ordinary differential equations in this paper, our methodology can naturally be extended to systems of partial differential equations that one wishes to reconstruct from spatiotemporal data. This would involve constructing function libraries (encoded by the matrix $\bg\Theta(\b X))$ that include all relevant partial derivatives, which can enlarge the column space of $\bg\Theta(\b X)$ substantially. This is a setting in which regularization is particularly necessary, as demonstrated in other PDE-focused studies, and where our methods can thus have great potential. Another avenue in which our methods can be extended could be the use of multivariate regression methods for vector-valued outcomes. In the context of dynamical system recovery, we typically consider coupled systems of differential equations in more than one dimension. Joint estimation of each equation in the system, as opposed to reconstructing each equation separately, may improve the accuracy of the overall reconstruction as well as reveal key interdependencies amongst the system's components. 

\section{Acknowledgements}

SV, MW and SB acknowledge partial support from the NIH award R01GM135926. In addition, SB also
acknowledges partial support from NSF awards DMS-1812128, DMS-2210675, CAREER DMS-2239102 and
NIH award R21NS120227.


\appendix

\section{Additional simulations: Van der Pol oscillator} \label{appsec:VdP}

\begin{figure}[H]
\includegraphics[width=\textwidth]{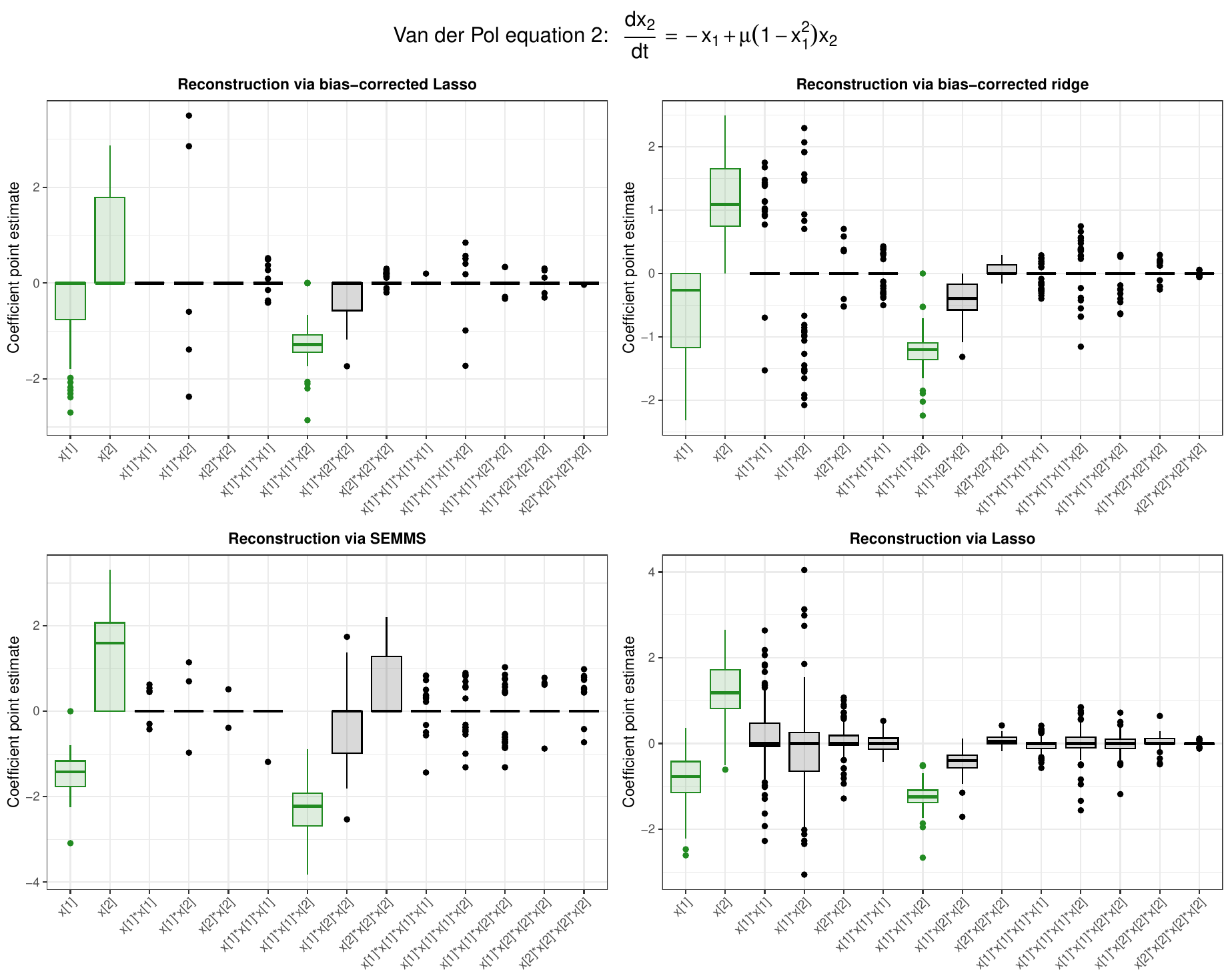}
\label{fig:boxplots}
\caption{Boxplots of the coefficients of each polynomial term up to fourth degree that could potentially be included in the reconstruction of the Van der Pol oscillator with $\mu=2$. These distributions were obtained by adding i.i.d. $N(0,0.25m)$ noise to the numerical solution of the system 100 times, where $m=\max_{i,j}|\b X_{ij}|$, and estimating the coefficients of each term on each of these 100 noisy datasets. Correct terms, i.e. those that do appear in the equation, are highlighted in green. We observe that the boxplots for incorrect terms tend to be much more tightly concentrated around 0 under the bias-corrected regression estimators as well as SEMMS, and that there are fewer outliers, compared to the Lasso.}
\end{figure}

\begin{figure}[H]
\includegraphics[width=\textwidth]{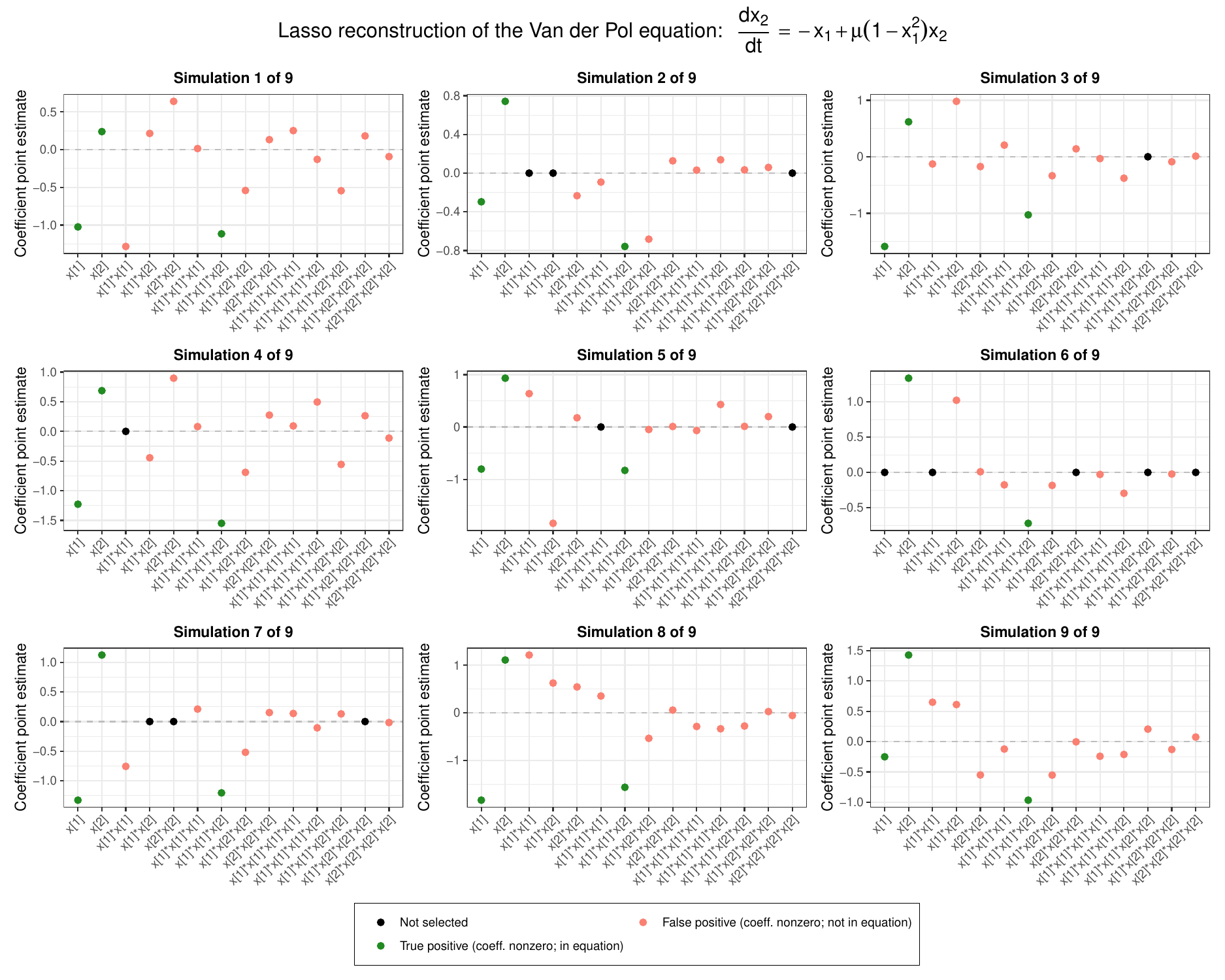}
\label{fig:LassoPlots}
\caption{Lasso coefficient estimates for each polynomial term that could be included in the reconstruction of the Van der Pol oscillator with $\mu=2$ over nine simulations of noisy data from this ODE system. We observe that while the correct terms consistently have non-zero coefficients, the coefficients of the incorrect terms are highly variable and often non-zero.}
\end{figure}

\section{Additional simulations: 2D spiral equation} \label{appsec:spiral}

\begin{figure}[H]
\centering\includegraphics[width=0.85\textwidth]{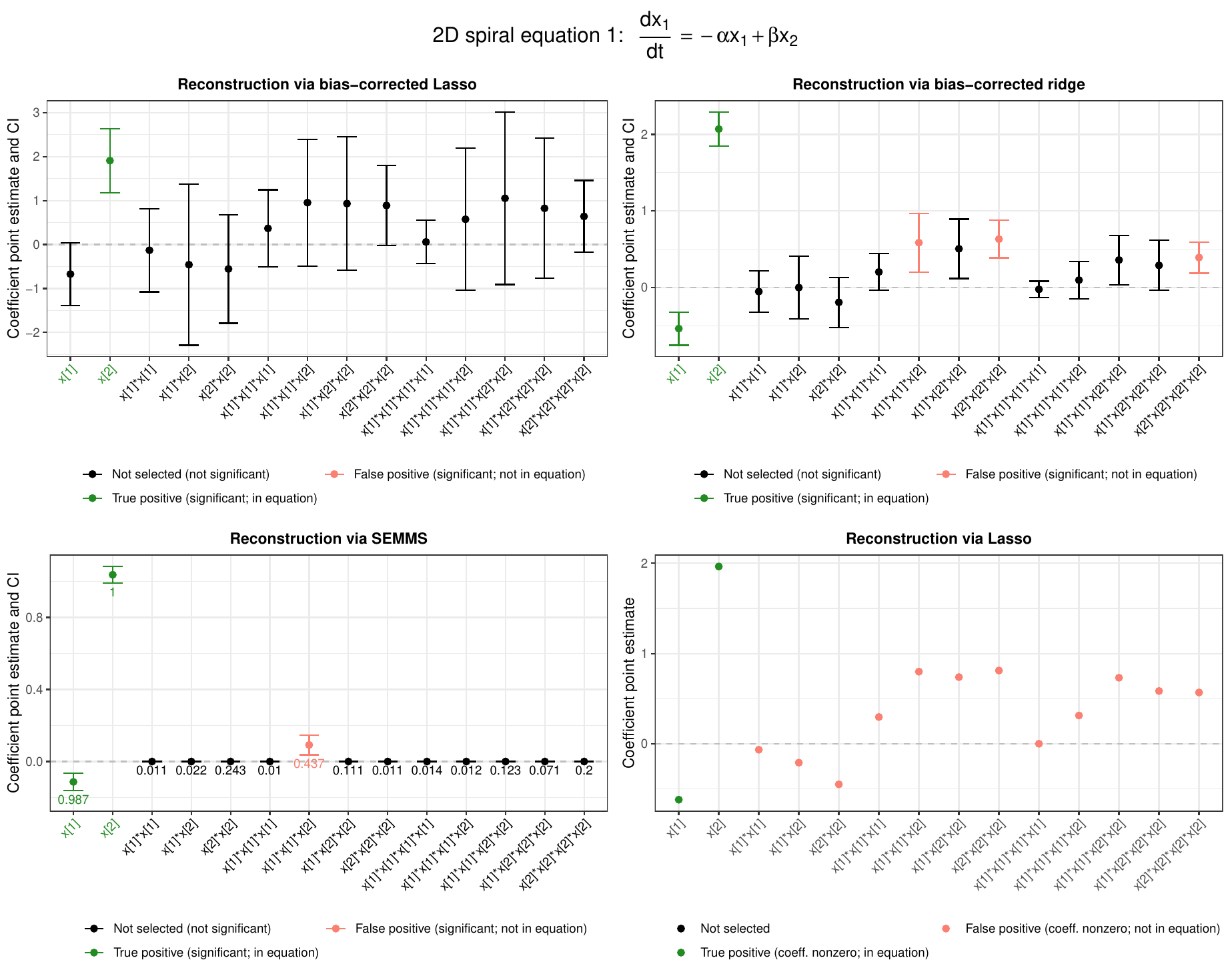}	
\caption{Estimates and error bars for the coefficients of each polynomial term up to fourth degree that could potentially be included in the reconstruction of a two-dimensional spiral equation, $\frac{\dv x_1}{\dv t}=-\alpha x_1 + \beta x_2$ with $\alpha =1/3$ and $\beta=3$. As in Figure \ref{fig:VdPcoeffs}, polynomial terms are included in the reconstructed equation if: they are statistically significant with a $p$-value $< 0.05$ under the bias-corrected Lasso (top left) or ridge (top right) estimators; they are selected by SEMMS (bottom left) based on their posterior probabilities of having non-zero coefficients; or if their Lasso coefficients (bottom right) are non-zero. As with the Van der Pol equation in Section \ref{sec:results}, the number of false positive terms is reduced substantially when using the bias-corrected Lasso, ridge, or SEMMS methods compared to the Lasso.}
\label{fig:spiralCoeffs}
\end{figure}

\begin{figure}[H]
\centering\includegraphics[width=\textwidth]{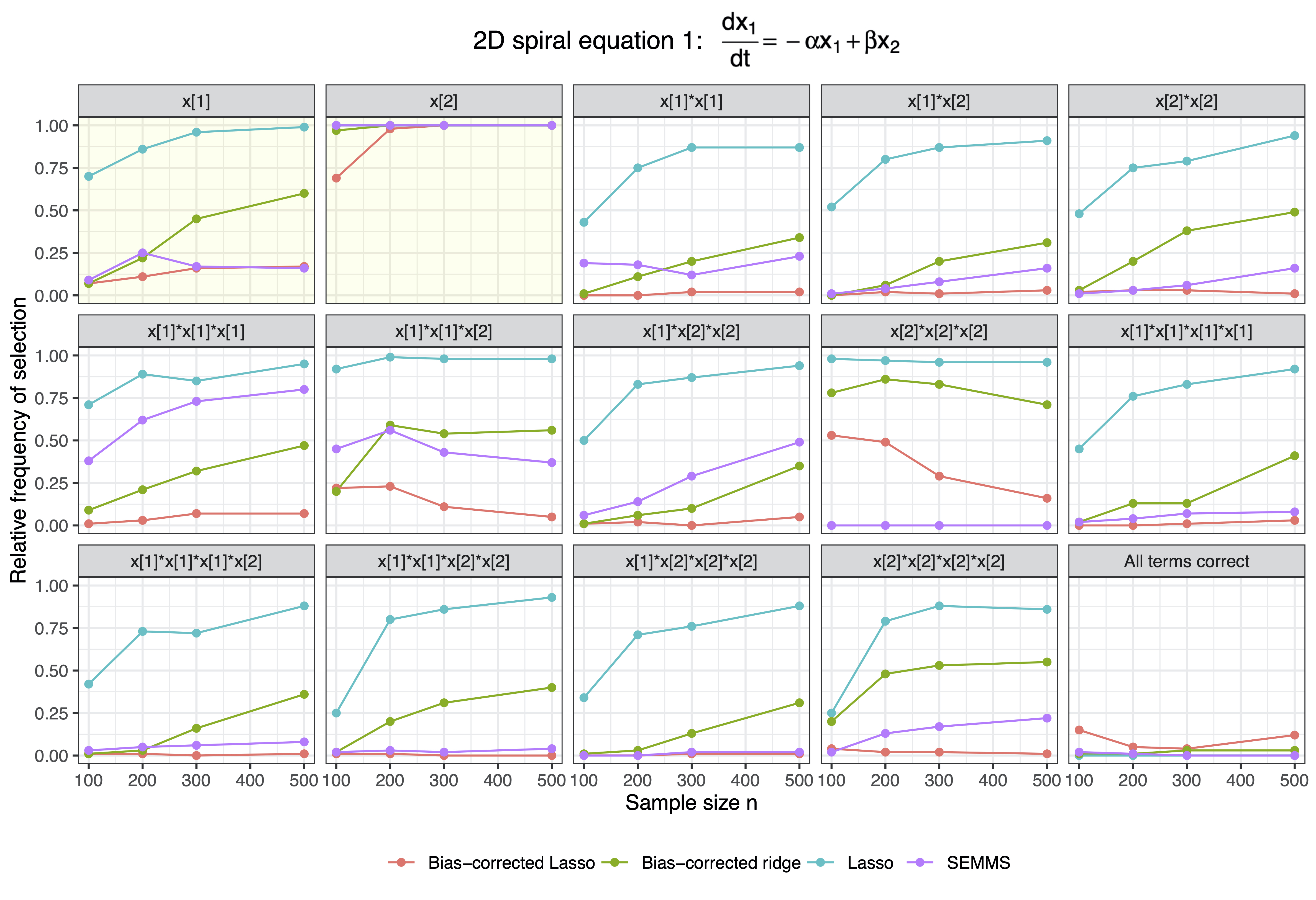}
\caption{Empirical probabilities of selection for each polynomial term up to fourth degree that could potentially be included in the reconstruction of the 2D spiral equation, $\frac{\dv x_1}{\dv t}=-\alpha x_1 + \beta x_2$  with $\alpha =1/3$ and $\beta=3$, for varying \textit{sample sizes}. The sample size $n$ refers to the number of time steps taken between $t = 0$ and $t = 20$ when numerically solving the spiral equation via RK4 (sample size increases with decreasing step size $h$).   Observe that the standard Lasso frequently selects incorrect terms even at larger sample sizes, whereas the bias-corrected Lasso and ridge methods as well as SEMMS do not pick up these terms nearly as often. The bottom right plot depicts the empirical probabilities of only these three terms appearing in the recovered equation, i.e. it indicates how often the correct equation was exactly identified. } 
\label{fig:spiralSampleSize}	
\end{figure}

\begin{figure}[H]
\centering\includegraphics[width=\textwidth]{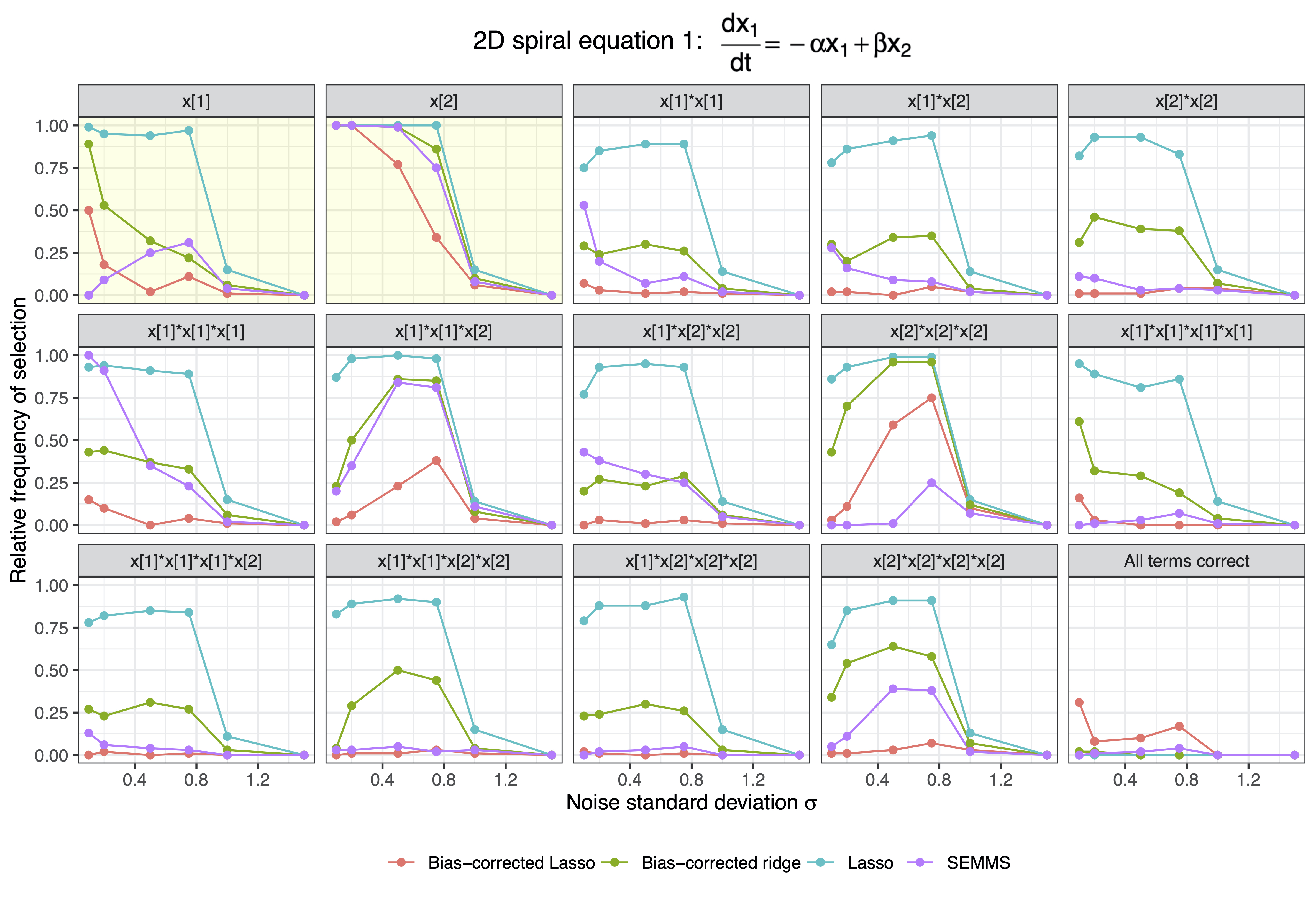}
\caption{Empirical probabilities of selection for each polynomial term up to fourth degree that could potentially be included in the reconstruction of the 2D spiral equation, $\frac{\dv x_1}{\dv t}=-\alpha x_1 + \beta x_2$ with $\alpha =1/3$ and $\beta=3$, for varying \textit{noise levels}. The noise level $\sigma$ refers to the standard deviation of the i.i.d. $N(0,\sigma)$ random variates that are added to the numerical solution of the 2D spiral equation obtained via RK4.}
\label{fig:spiralNoise}		
\end{figure}

In this section, we consider another two-dimensional ODE system and present simulation results similar to those in Section \ref{sec:results}. This ODE system, which we call the \textit{spiral equations} due to the shape of their plotted numerical solutions, is given by:
\begin{align}
\frac{\dv x_1}{\dv t} &= -\alpha x_1 + \beta x_2 \label{eq:spiral1} \\
\frac{\dv x_2}{\dv t} &= -\alpha x_2 - \beta x_2 \label{eq:spiral2}
\end{align}
where $\alpha$, $\beta \in \R$. As in Section \ref{sec:results}, we generate data from this system with $\alpha=1/3$ and $\beta=3$ using the RK4 method with an initial condition of $\b x_0=[x_1(t_0)~~ x_2(t_0)] = [2,0]$ and step size of 0.05 over the time domain $t\in[0,20]$, resulting in $n=400$ time steps. We again add i.i.d. $N(0,0.25m)$ noise, where $m$ is the largest value in magnitude in the state matrix $\b X$. We consider a library of possible polynomial terms up to fourth degree, and aim to correctly identify the terms $x_1$ and $x_2$ for both equations \eqref{eq:spiral1} and \eqref{eq:spiral2}. In Figure \ref{fig:spiralCoeffs} we display the coefficient point estimates and their confidence intervals, and in Figures \ref{fig:spiralSampleSize} and \ref{fig:spiralNoise} we present how often the correct terms are identified under varying sample sizes and noise levels.

\bibliography{References}
\bibliographystyle{abbrvnat}


\end{document}